\documentclass[aps,prb,twocolumn,showpacs,amsmath,amssymb,unsortedaddress]{revtex4-1}
\usepackage{graphicx}
\usepackage{longtable}
\usepackage{ulem}
\usepackage{color}

\begin{document}
\title{Design and construction of a compact rotary substrate heater for deposition systems}

\author{Israel Perez} 
\email[Contact Author: ]{cooguion@yahoo.com}
\affiliation{National Council of Science and Technology (CONACYT)-Department of Physics and Mathematics, Institute of Engineering and Technology, Universidad Aut\'onoma de Ciudad Ju\'arez, Av. del Charro 450 Col. Romero Partido, C.P. 32310, Ju\'arez, Chihuahua, M\'exico}
\author{Tareik Netro}%
\affiliation{Department of Physics and Mathematics, Institute of Engineering and Technology, Universidad Aut\'onoma de Ciudad Ju\'arez, Av. del Charro 450 Col. Romero Partido, C.P. 32310, Ju\'arez, Chihuahua, M\'exico}%
\author{Mario Vazquez}%
\affiliation{Department of Physics and Mathematics, Institute of Engineering and Technology, Universidad Aut\'onoma de Ciudad Ju\'arez, Av. del Charro 450 Col. Romero Partido, C.P. 32310, Ju\'arez, Chihuahua, M\'exico}%
\author{Jos\'e  Elizalde}%
\affiliation{Department of Physics and Mathematics, Institute of Engineering and Technology, Universidad Aut\'onoma de Ciudad Ju\'arez, Av. del Charro 450 Col. Romero Partido, C.P. 32310, Ju\'arez, Chihuahua, M\'exico}%
\date{\today}

\begin{abstract}
We have designed and constructed a compact rotary substrate heater for the temperature range from 25 $^\circ$C to 700 $^\circ$C. The heater can be implemented in any deposition system where crystalline samples are needed. Its main function is to provide a heat treatment in situ during film growth. The temperature is monitored and controlled by a temperature controller coupled to a type K thermocouple. A heater case was designed to host a resistive element and at the same time to allow the substrate holder to freely rotate. Rotation is crucial not only for film homogeneity during deposition but also for the elimination of temperature gradients on the substrate holder. To tolerate oxidizing and corrosive environments, the instrument was made of stainless steel which also works as ``coolant" taking advantage of heat dissipation. The instrument performs well for long periods of time with stable temperatures. We hope that this project is useful for laboratories wishing to have a compact rotary heater that meets the requirements for crystal growth and film homogeneity.
\end{abstract}
\maketitle
\section{Introduction}
\label{intro}
A substrate heater is one of the most important components in depositions systems. Its function is to give enough energy to adatoms on the substrate for their adsorption and therefore for the formation of crystalline structures. Nowadays, substrate heaters can be found in many deposition techniques such as DC and radio frequency (RF) magnetron sputtering, electron beam evaporation (EBE), plasma assisted deposition (PAD),  pulsed laser deposition (PLD), atomic layer deposition (ALD), molecular beam epitaxy (MBE), and several chemical vapour deposition (CVD) techniques such as laser chemical vapour deposition\cite{eatanassova95a,hshinriki91a,skamiyama93a,gqlo93a,ykuo92a,ndonkov11a,skr88,tjj94a,hop99a, kssh06a,dmm10a,sb05a,gb10a}. 

Heaters have been developed for specific purposes based on needs and several heating methodologies. The most commonly used technologies employ filaments and heating lamps \cite{pls93a,tej94a,amh10a}. Heating lamps are used in deposition equipment where the instrumentation is not at risk of being damaged by the emitted radiation. However, in some deposition systems such as sputtering this radiation can affect sensible components, namely: magnetrons and targets themselves. Some systems also use quartz crystal monitors (QCM) whose measurement depends not only on pressure but also on temperature, and therefore the operation of this instrumentation is largely determined by heat. To circumvent this problem a localized source of heat is recommendable and this is done with a resistive element or filament. The introduction of a filament into the vacuum chamber poses a problem if not handle with care. The materials the filaments are made of can also evaporate and contaminate the source material (e.g. evaporants, targets, etc.), the film under growth, and the vacuum chamber. If filaments, however, are well shielded, this effect can be negligible or even eliminated. The most common resistive elements found in the market are tungsten wire, Khantal-A1 wire, and nichrome wire. The melting point of tungsten is one of the highest among metals (3420 $^\circ$C) as such, tungsten filaments are widely utilized in high temperature furnaces. A disadvantage of tungsten wire is that is brittle and can easily break if not handle with care. Moreover, tungsten wire is expensive in comparison to the other two. Khantal-A1 and nichrome wires are more malleable and have lower fusion points ($\sim 1500\,^\circ$C and $\sim 1400\,^\circ$C, respectively); so they are used in middle temperature furnaces. Despite of their usefulness, these materials are highly specialized and not easy to find in some regions; so they must be ordered from far places increasing the overall cost.

On the other hand, in deposition techniques where the substrate is static, gradients of heat and/or of atomic concentrations can develop on the substrate as deposition evolves, exposing the film to atomic density inhomogeneities and mechanical stresses. These problems can be undermined using a rotary heater. Its implementation however, poses an additional technical problem in its design, increasing its cost, specially for commercial heaters. In the past, several developers have reported the construction of heaters for deposition systems. For instance, P. L. Stewart et al. developed a substrate heater based on a 900 W tungsten lamp \cite{pls93a}. The lamp is large, radiates heat in all directions and the maximum substrate temperature is 450 $^\circ$C which is relatively low for the crystallization of many other compounds. This kind of heater is limited to some type of vacuum chambers where instrumentation is not affected by radiation, e.g., EBE. On the other hand, T. E. Jones et al. constructed a compact and resistive-based heater capable of working in oxygen atmospheres and reaching temperatures close to 800 $^\circ$C. However, thermal isolation is poor and radiation is also spread in all directions\cite{tej94a}. Besides, both of these designers did not include a rotation mechanism. In this sense, A. M. Herrera et al. also fabricated a compact rotary heater with good thermal isolation, reaching a maximum temperature of 800 $^\circ$C. To isolate the filament, they used a ceramic case with a flow of coolant \cite{amh10a}. Although this is a good idea, the inclusion of the cooling systems increases design complexity and thus overall cost. As seen in their images their heater looks robust (no information on dimensions is given) and might not be implemented in chambers with limited space. Furthermore, all of these heaters rely on DC powers supplies with low voltages and high current intensities. These currents can easily generate static magnetic fields around the filament that can interfere with charged particles inside the chamber (electrons, argon cations, oxygen anions, etc.). This represents an additional unwanted parameter that can produce preferential deposition affecting the homogeneity and the crystallinity of the films.

To address some of these issues, in this work we propose the design and construction of a compact rotary substrate heater for a deposition system. The heater was implemented in a RF sputtering system and is capable of reaching a maximum temperature of 700 $^\circ$C based on a  commercial filament driven by a variable AC power supply (VARIAC). The temperature is controlled by varying the power and monitored by a temperature controller coupled to a type K thermocouple. The usage of a VARIAC and a commercial filament, makes this heater easy to implement in any deposition system without sacrificing temperature accuracy and stability as well as film homogeneity and crystallinity.

\section{Design and construction}
\subsection{Design considerations}

For the design of the instrument we considered mainly the maximum temperature, system geometry, requirements of the vacuum system, and versatility for the growth of different compositional films. The most important parameters are system geometry and the maximum temperature; once these are chosen the rest of the considerations must be tailored to fit these aspects. 

Many chemical compounds, synthesized by the so-called solid state reaction require temperatures around 1000 $^\circ$C. However, these compounds can crystallize in vacuum under controlled conditions of pressure at temperatures below 700 $^\circ$C. We thus think that this is a reasonable maximum temperature for our heater, so we can grow a myriad of composition films. 

As a heating method we have chosen heating by radiation using a resistive element. The main reason is because the radiation given off by the filament can be encapsulated in a small compartment or case.  Also filaments occupy less space than lamps and can be shaped as desired. In this kind of set up, the filament is used to heat the substrate holder by radiation and in turn the substrate holder heats the substrate by conduction. Heating by radiation in vacuum is challenging because heat is heavily lost in a few millimetres away from the source and due to the low pressure heating by convection is inefficient. For instance, if tungsten filament of 24 mm length had a temperature on its surface of 2350 $^\circ$C, the temperature 1 cm away would fall about 1500 $^\circ$C. For this reason the filament must be placed as close as possible to the substrate holder \cite{rrk98a}. To enclose the radiation we have decided to encapsulate the filament in a cylindrical metallic case. The cap of this cylinder being the sample holder placed at about 300 $\mu$m away from filament. This case also helps to dissipate the heat along the structure of the heater  avoiding the implementation of a cooling system. 

 To attain temperatures around 700 $^\circ$C high electric currents are required. Traditionally, designers used DC power supplies, however, these supplies work with voltages of less than 60 V and thus, for a given power, current intensities for the filament can be as high as 30 A. A high current generates a static magnetic field that can be strong enough (several gauss) to deviate ions and electrons inside the chamber, producing a preferential orientation of adatoms on the film and at the end affecting the crystal structure. Moreover, in chambers equipped with low energy and reflexion high energy electron diffraction (LEED and RHEED, respectively), the magnetic field can heavily deviate the electron beam from its original trajectory and create distorted patterns on the fluorescent screen. This is an important issued that heater designers usually overlooked. So, to lessen the effect of the magnetic field we have resolved using a VARIAC. With this power supply we feed the filament and control the power delivered by either sending pulses of current or varying the applied voltage, taking advantage of the fact that filaments generally obey Ohm's law even at high temperatures, as high as 2000 $^\circ$C. 

Our next problem is crucial during film growth. As is well known, in most techniques (such as RF sputtering, PLD, EBE, etc.) atoms are ejected from the crucible or target and spread out in all directions forming the so-called deposition cone. Generally, this cone has an inhomogenous atomic density and this can largely affect the homogeneity of the film. To solve this problem, substrate holders are normally rotated at low speeds (less than 2 rpm). Rotating the substrate holder also helps to substantially minimize temperature gradients caused by the geometry of the heating element. With this in mind, we implemented a rotary sample holder driven by a motor.

\begin{figure}[t!]
\centering
\includegraphics[width=8cm]{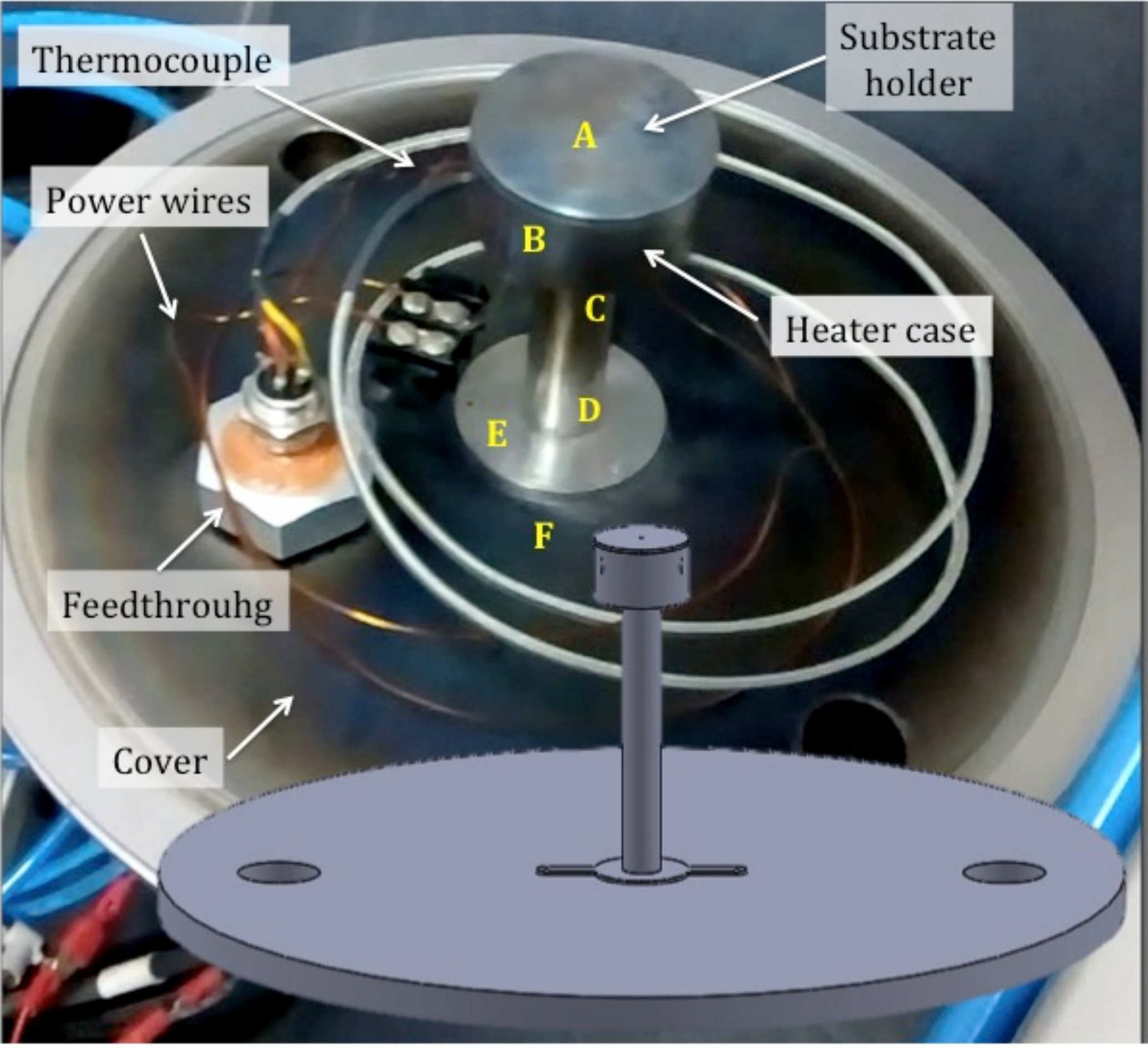}
\caption{Heater mounted on cover. The drawing is the design and the picture shows the actual heater}
\label{base}
\end{figure}

The following challenge is to monitor the substrate temperature. An option to do this is to use an infrared pyrometer. The pyrometer can be placed outside the chamber to measure the radiated heat given off by the substrate. However, this, of course, implies that there is an available viewport and sometimes this is not the case. More important is the fact that, since many different kind of substrates and films are grown each deposition session, the pyrometer needs to be calibrated for each case; turning this method a tedious procedure. Moreover if economy is an important issue the use of a pyrometer is not a good option.  We thus have chosen to use a thermocouple and hold it as close as possible to the sample holder (see figure \ref{base}).

\subsection{Design and Construction}
\begin{figure}[b!]
\centering
\includegraphics[width=6cm]{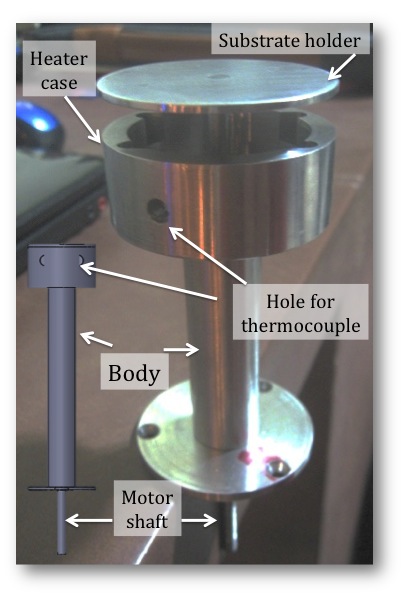}
\caption{Heater, showing the motor shaft, body, heater case and substrate holder. The inset shows the design}
\label{shaft}
\end{figure}

\begin{figure}[t!]
\centering
\includegraphics[width=9.5cm]{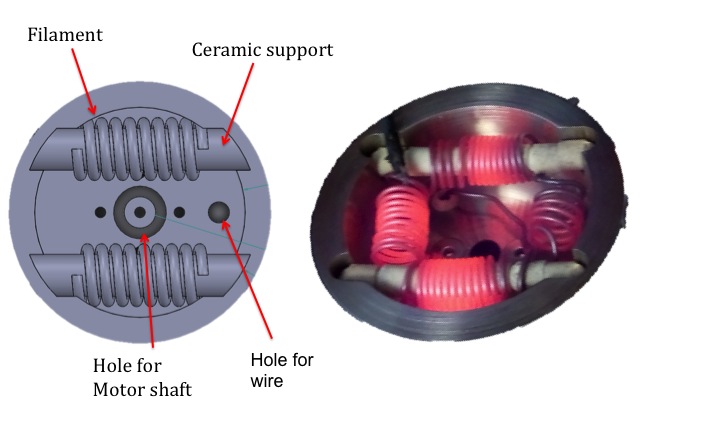}
\caption{Interior of the heater showing the filament with the supports. Left is the design and right is the actual heater working}
\label{horno}
\end{figure}

Taking into account the above considerations we proceeded to design and construct the heater. In figures \ref{base}-\ref{horno} we show the design of the heater along with the actual heater. The heater can be scalable to any size and, according to needs, implemented in any deposition system. The instrument is mainly composed of five parts (see figure \ref{shaft}), from bottom to top we have: a motor shaft, a bottom base, a heater body, a heater case, and the sample holder. 

The bottom base is used to firmly hold the rest of the heater components to the cover plate of the vacuum chamber. It is a small cylinder, with a diameter of 60 mm and a thickness of 3 mm which is attached to the cover plate by three bolts. At the center, there is a 5 mm hole to let pass the motor shaft. 

The heater body is a shallow cylinder with a 5 mm hole drilled along the axis to host the motor shaft. The cylinder dimensions are 75 mm $\times$ 20 mm and it is attached to the heater case at the upper end and to the bottom base at the lower end. On the other hand, the heater case acts as a heat shield and filament host. It is also a shallow cylinder with an external diameter of 50 mm and an external height of 20 mm. The thickness of the wall is 5 mm, leaving an interior diameter of 40 mm. At the center there is a 5 mm hole to let pass the motor shaft. Two 5 mm holes were drilled on the bottom of the case to feed the filament. Another 5 mm semi-spherical hole was drilled on the side of the heater case to host the thermocouple. Two pairs of grooves were also drilled on the interior diameter (see figures \ref{shaft} and \ref{horno}). These grooves hold two metallic bars holding the ceramic beads where the filament is wound. The groove length is 44 mm with a thickness of 5 mm. With respect to the heater case, they are placed at a height of 7 mm so that the filament remains about 300 $\mu$m away from the sample holder. The function of this clearance is two-fold: first, avoids undesirable mechanical friction between the sample holder and the heater case, allowing a free rotation of the sample holder; and second, traps filament radiation as much as possible.

The sample holder is a flat cylinder with a diameter of 50 mm and a thickness of 1.5 mm attached to the upper end of the motor shaft. The motor shaft has a diameter of 4.9 mm and a length of 150 mm. On the upper end, the sample holder is welded to the motor shaft, then the motor shaft is inserted in the body of the heater, passing through the center of the heater case (see also figure \ref{horno}). On the lower end, the motor shaft has a thread for coupling with the rotary feedthrough mounted on a central chamber port (flange KF25) located on the opposite site of the cover plate (see figure \ref{motor}). The rotary feedthrough was acquired from Intercovamex manufacturer and seals using a plastic o-ring surrounding  the motor shaft that has a limiting lock to leave the clearance mentioned above. Since the substrate holder must rotate freely, a low power motor is required (less than 10 W). The motor is held on a stainless steel square plate (4 mm thick and area of 10 cm $\times$ 10 cm) that in turn is held to the port by four long bolts. 

If service is required, for instance to change the filament, the substrate holder is just rotated counter clockwise and pull out, exposing the filament and filament supports for easy maintenance.

Since we are interested in measuring the substrate temperature, it would be ideal to place the thermocouple on the substrate holder. However, if this were the case, the thermocouple would rotate and the cable would get entangled around the heater body. To avoid this problem, we inserted the thermocouple in the hole drilled on the side of the heater case (see figure \ref{shaft}) and glued it with silver paint for good heat conduction. In this way the thermocouple does not move, although the temperature is not the temperature of the substrate holder. Later we will deal with this issue during calibration. The power wires and the thermocouple cable were connected to the power supply and the temperature controller, respectively, through an electrical feedthrough acquired, as well, from Intercovamex  (see figure \ref{base}). 

\subsection{Instrumentation and Materials Used}
For the heating element we used a 10 AWG Boccherini resistance (other brands such as Coflex and FOSET are typically found in the market) commonly used for electric showers and purchased from a known hardware store. As such the resistance is much less expensive than Khantal-A1 and nichrome wires. The filament is an alloy of nichrome-copper which is pretty malleable and designed to tolerate rough environments, such as working at atmospheric pressure and atmospheric conditions. The material has an chromium oxide layer that protects the filament from corrosive environments. Filaments have a low electrical resistance, typically less than 10 $\Omega$, the resistance depending on its length, for this reason they require high currents to generate enough heat. The electric resistance of our filament as measured with a multimeter is 2.9 $\Omega$ at 27 $^\circ$C. Supports for the filament are composed of metallic bars covered with commercial alumina beads (also acquired from a local retailer) whose melting point is close to 2000 $^\circ$C. Alumina beads were mainly used in our system as electrical and thermal insulators against metallic parts.

The metallic components of the heater were machined of stainless steel grade 306. According to manufacturer the fusion point is 1400 $^\circ$C. Stainless steel is one of the preferred materials due to its resistance to corrosion, withstanding a wide range of temperatures without undergoing deformation. The thermal conductivity ranges from 15 W/m$\cdot$K to 22 W/m$\cdot$K at 100 $^\circ$C and 500 $^\circ$C, respectively, which is less than 10 times that of copper (300 W/m$\cdot$K). Because of this we expect not to have a strong heat diffusion, so the temperature in the substrate holder remains stable as thermodynamic equilibrium is reached for a given applied power. Since the heater case is mounted on the body cylinder, this in turn is attached to the bottom base and the latter to the cover, heat diffusion is mild, thus avoiding overheating the vacuum chamber by heat conduction.

\begin{figure}[t!]
\centering
\includegraphics[width=8cm]{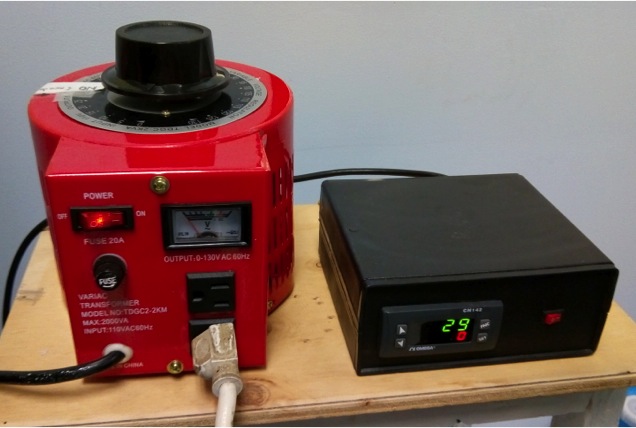}
\caption{VARIAC and temperature controller}
\label{tempcont}
\end{figure}
To measure the temperature we used a temperature controller Omega CN142 coupled to a type K thermocouple both purchased from Omega manufacturer (see figures \ref{base} and \ref{tempcont}). This controller is based on the proportional integral-derivative (PID) technology and has a relay output that allows to control the power delivered to the filament. The thermocouple was attached to the heater case on one side as close as possible to the heating element (see figure \ref{shaft} for the exact position) and then connected to the temperature controller. 

\begin{figure}[t!]
\centering
\includegraphics[width=5cm]{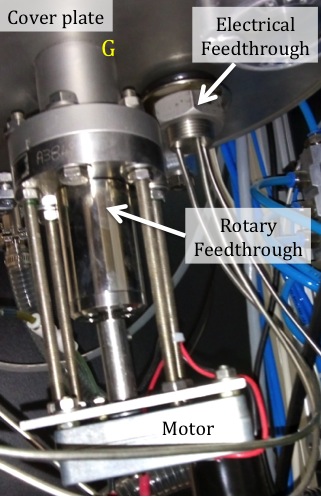}
\caption{Opposite side of the cover plate. Electric motor and rotary feedthrough used for rotating the substrate holder. The point G indicates an additional point were temperature was measured.}
\label{motor}
\end{figure}

We acquired from Acomee manufacturer an AC variable power supply (VARIAC) with a variable output voltage, 0 V-160 V, giving a maximum power of 2 KVA (see figure \ref{tempcont}). To avoid voltage surge transients, the VARIAC was plugged to an uninterruptible power system (UPS) that keeps the output voltage stable. This guarantees a stable output power. As for the motor, we have used a 12 VDC motor (see figure \ref{motor}). The motor speed is regulated varying the applied voltage coming from a variable DC power supply.

\section{Results}
\subsection{Initial tests}

\begin{figure}[t!]
\centering
\includegraphics[width=9cm]{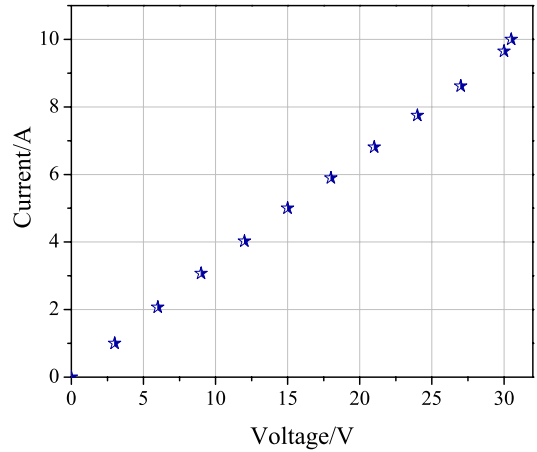}
\caption{Current-voltage characteristic of filament. The filament resistance remains constant and stable for the whole interval}
\label{VvsIcopy}
\end{figure}

To check the performance of the heater, we carried out several initial tests. The first test was to measure the current-voltage characteristics of the filament to ensure that the resistance is linear as a function of temperature. The results of these measurements are given in figure \ref{VvsIcopy}. Here we can clearly see that there is an Ohmic relation. By a linear fit we obtained a constant resistance of 3.0 $\Omega$ in close agreement with the value measured with the multimeter at room temperature. This linear relation indicates the stability of the resistance as the voltage and temperature increase.

\begin{figure}[b!]
\centering
\includegraphics[width=9cm]{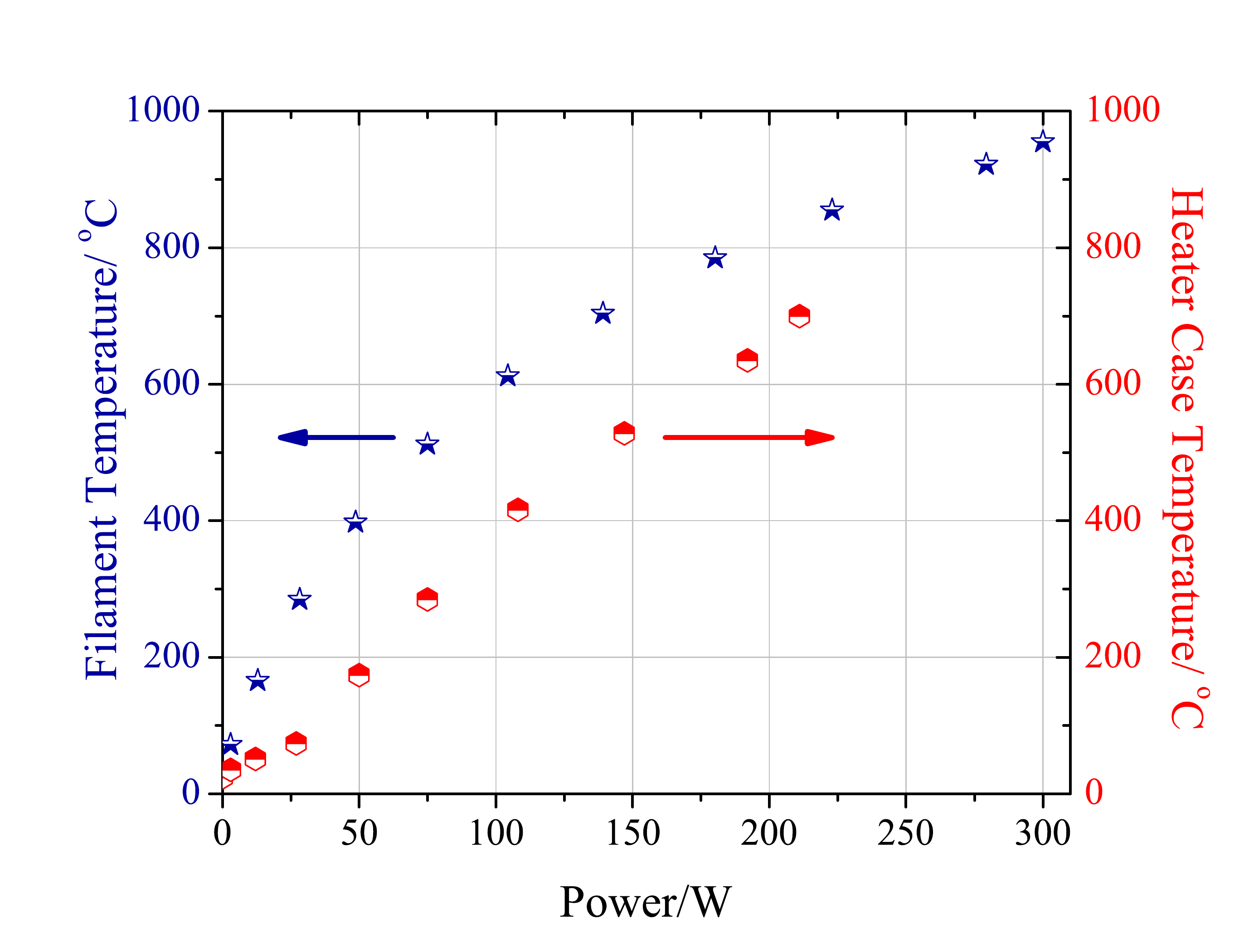}
\caption{Power dependence of temperature. Left scale is the temperature as measured on the filament. Right scale is the temperature as measured on the heater case.}
\label{PvsFilaTemp}
\end{figure}
The next test gave us information of the power dependence of the filament temperature. The power was computed as $P=I^2R$ and the results are given in figure \ref{PvsFilaTemp}, left scale. The temperature on the filament was measured with an additional thermocouple (plugged to a multimeter Steren, MUL-605) for the interval from room temperature up to 955 $^\circ$C. We can observe a nonlinear relationship between power and filament temperature. These results are typical in filament heaters \cite{tej94a} and were taken as reference for calibration purposes.

\subsection{Performance}
As we mentioned above, the thermocouple is not placed on the substrate holder but on the side of heater case as close as possible to the substrate holder. The key idea is that the temperature at this point is representative of the temperature on the substrate holder. However, it is clear that due to heat dissipation, the temperature on these two points differs and a correction on the temperature controller must be included. To assess heat dissipation all over the chamber we measured the temperature in several points with the additional thermocouple. The points labelled from A to F in figure \ref{base} indicate the exact points where the temperature was measured when the temperature on the sample holder (point A) was 700 $^\circ$C after a 100 min plateau. The results are (700, 545, 179, 110, 71, 55) $^\circ$C, respectively. Additionally, we also measured the temperature on the opposite side of the cover plate (point G in figure \ref{motor}) and found a temperature of 41 $^\circ$C. The rest of the walls of the chamber remained at room temperature. Accordingly, we can see that heat diffusion by conduction in the chamber is poor due to the thermal conductivity of stainless steel. It is worth noting that for a power of 200 W the maximum temperature in the filament is 830 $^\circ$C whilst for this same power the temperature on point B is 545 $^\circ$C, thus losing 285 $^\circ$C by radiation from the filament to the heater case.
\begin{figure}[t!]
\centering
\includegraphics[width=8.5cm]{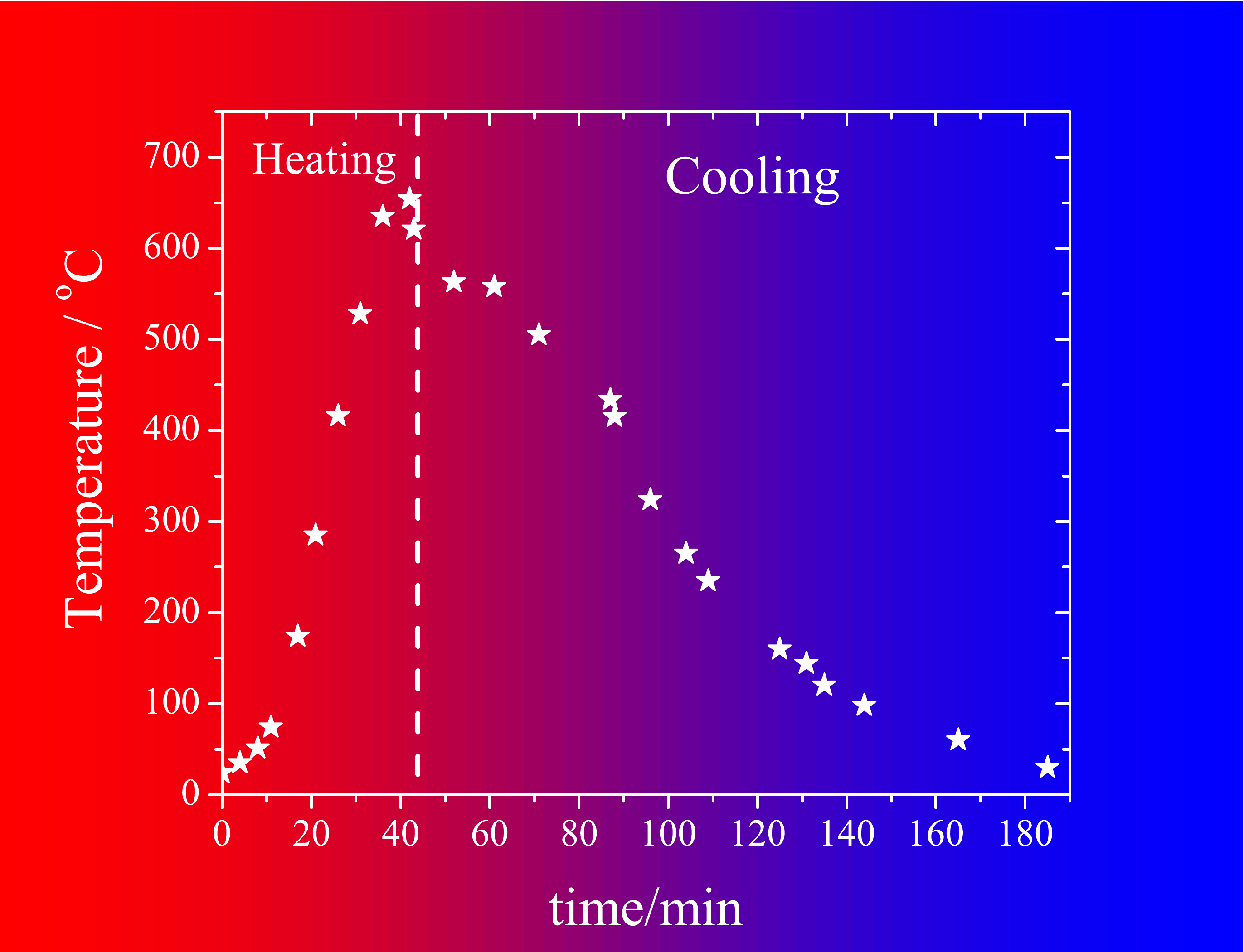}
\caption{Heating and cooling ramps. Vertical line defines the inflection point}
\label{CoolingHeating}
\end{figure}

Bearing these results in mind, we compensated the difference in temperature between the temperature on the sample holder and on the heater case, so the thermocouple placed on the heater case reads the actual temperature on the sample holder. This was done by setting an offset temperature in the temperature controller. After calibration, we measured the temperature up to 700$^\circ$C with the thermocouple fixed to the heater case (thus representing the actual temperature on the substrate holder) as a function of power. The results are shown in figure \ref{PvsFilaTemp}, right scale. We can see that the relationship between temperature and power is close to linear.

Finally, we measured the heating and cooling ramps for a maximum temperature of 654 $^\circ$C. The results are displayed in figure \ref{CoolingHeating}. The vertical line divides the heating ramp from the cooling ramp. As we can see it takes about 40 min to reach the maximum temperature, and just 25 min to reach 400 $^\circ$C which are usual times for heaters in vacuum chambers. In fact, this a typical time for pumping down or venting a vacuum chamber. Since we have not implemented a cooling system, the heater cools down by heat interchange with its surroundings. The cooling ramp takes much longer, about two hours to go from 654 $^\circ$C to room temperature and about 1 hour to go from 400 $^\circ$C to room temperature. This is indeed a disadvantage with respect to heaters with cooling systems implemented\cite{amh10a}. Yet, annealing a sample and/or venting the chamber helps to catalyze the cooling process and the time waiting for cooling is less than an hour even for the maximum temperature.

\section{Conclusions}
We have designed and constructed a compact rotary heater which was implemented in a RF magnetron sputtering system. The heating element is a filament fed with an AC power supply. The heater is capable of rotating and maintaining a stable temperature of 700 $^\circ$C for long periods of time providing thermal energy for film crystallization. Rotation contributes to eliminate unwanted temperature and deposition gradients, favouring the growth of homogenous films. Further, if the filament needs to be replaced, the heater was designed for an easy maintenance. Lastly, we have shown that even without cooling system the heater cools down in a reasonable time. We conclude that the heater constructed here fulfills the requirements for most deposition systems.

\section*{Acknowledgements}
 The authors gratefully acknowledge the support from the National Council of Science and Technology (CONACYT) Mexico, the program C\'atedras CONACYT through project 3035 and to project PIVA UACJ number 334-18-12. We are also indebted to the anonymous reviewer for valuable comments that greatly improved the quality of this work.


\begin{thebibliography}{99}

\bibitem{eatanassova95a} E. Atanassova, T. Dimitrova, J. Koprinarova, AES and XPS study of thin RF-sputtered Ta$_2$O$_5$ layers, Appl. Surf. Sci. 84, 193-202 (1995)

\bibitem{hshinriki91a} H. Shinriki, M. Nakata, UV-O$_3$ and Dry-O$_2$: Two-step Annealed Chemical Vapor-Deposited Ta$_2$O$_5$ Films for Storage Dielectrics of 64-Mb DRAMÕs, IEEE Trans. Electron Devices ED. 38, 455-462 (1991)

\bibitem{skamiyama93a} S. Kamiyama, P.-Y. Lesaicherre, H. Suzuki, A. Sakai, I. Nishiyama, A. Ishitani, J. Electrochem. Soc. 140, 1617 (1993)

\bibitem{gqlo93a}  G.Q. Lo, D.L. Kwong, S. Lee, Appl. Phys. Lett. 62, 973, (1993)

\bibitem{ykuo92a} Y. Kuo, Reactive Ion Etching of Sputtered Deposited Tantalum Oxide and its Etch Selectivity to Tantalum, J. Electrochem. Soc. 139, 579-583, (1992)

\bibitem{ndonkov11a} N. Donkov, A. Zykova, V. Safonov, R. Rogowska, J. Smolik, Tantalum Pentoxide Ceramic Coatings Deposition on Ti4A16V Substrates for Biomedical  Applications, Problems At. Sci. Technol. Plasma Physics Series 17, 131-133, (2011)

\bibitem{skr88} S. K. Roy, Laser Chemical Vapour Deposition, Bull. Mater. Sci. 11, 129-135 (1988)

\bibitem{tjj94a} T. J. Jackson and S. B. Palmer, Oxide superconductor and magnetic metal thin film deposition by pulsed laser ablation: a review, J. Phys. D: Appl. Phys., 27, 1581-1594 (1994)

\bibitem{hop99a} H. O. Pierson, Handbook of Chemical Vapor Deposition, 2nd Edition, Noyes publications (1994)

\bibitem{kssh06a} K. S. Sree Harsha, Principles of Physical Vapour Deposition of Thin Films, Elsevier (2006) 

\bibitem{dmm10a} D. M. Mattox, Handbook of Physical Vapor Deposition (PVD) Processing, William Andrew (2010)

\bibitem{sb05a} S. Berg, T. Nyberg, Fundamental understanding and modeling of reactive
sputtering processes, Thin Solid Films, 476, 215-230 (2005)

\bibitem{gb10a} G. Bra\"uer, B. Szyszka, M. Vergo\"ohl, R. Bandorf, Magnetron sputtering-Milestones of 30 years, 84, 1354-1359 (2010)

\bibitem{pls93a} P.L. Swart, B.M. Lacquet y S. Reynecke, IEEE Trans. on Nucl. Sci. 40, 3 (1993)

\bibitem{tej94a} T.E. Jones, W.C. McGinnyis y J.S. Briggs, Rev. Sci. Instrum. 4, 65 (1994) 

\bibitem{amh10a} A. M\'arquez-Herrera, E. Hern\'andez-Rodr\'iguez, M.P. Cruz-J\'auregui, M. Zapata-Torresa y A. Zapata-Navarro, Calentador de sustratos compacto y de bajo costo para tratamiento t«ermico in situ de pel\'iculas delgadas depositadas por rf-sputtering, Rev. Mex. Fis., 56, 85-91, (2010)

\bibitem{rrk98a} Raghavendra Rao Kanchi and Naveen Kumar Uttarkar, Study of Heat Loss from Hot Tungsten Filament Bulb Using AT89C51 Based Data Acquisition System, Int. J. Appl. Phys. Math, 2, 194, (1998)

\bibitem{erc01a} Emerson R. Camargo and Masato Kakihana, Chemical Synthesis of Lithium Niobate Powders (LiNbO3) Prepared from Water-Soluble DL-Malic Acid Complexes, Chem. Mater., 13, 1905, (2001)

\end{thebibliography}
\end{document}